\documentclass[conference]{IEEEtran}
\usepackage{times}
\usepackage{epsfig}
\usepackage{graphicx}
\usepackage{amsmath}
\usepackage{amssymb}
\usepackage{booktabs}
\usepackage{algorithm}
\usepackage{algorithmic}
\usepackage{hyperref}

\title{Stock Price Prediction Using Triple Barrier Labeling and Raw OHLCV Data: Evidence from Korean Markets}

\author{Sungwoo Kang$^{a}$ (krml919@korea.ac.kr), Jong-Kook Kim$^b$ (jongkook@korea.ac.kr)\\
\small
$^a$ Department of Electrical and Computer Engineering, of Korea University Seoul 02841, Republic of Korea\\
$^b$ School of Electrical Engineering, of Korea University Seoul 02841, Republic of Korea
}

\begin{document}

\maketitle

\begin{abstract}
This paper demonstrates that deep learning models trained on raw OHLCV (open-high-low-close-volume) data can achieve comparable performance to traditional machine learning (ML) models using technical indicators for stock price prediction in Korean markets. While previous studies have emphasized the importance of technical indicators and feature engineering, we show that a simple LSTM network trained on raw OHLCV data alone can match the performance of sophisticated ML models that incorporate technical indicators. Using a dataset of Korean stocks from 2006 to 2024, we optimize the triple barrier labeling parameters to achieve balanced label proportions with a 29-day window and 9\% barriers. Our experiments reveal that LSTM networks achieve similar performance to traditional machine learning models like XGBoost, despite using only raw OHLCV data without any technical indicators. Furthermore, we identify that the optimal window size varies with model hidden size, with a configuration of window size 100 and hidden size 8 yielding the best performance. Additionally, our results confirm that using full OHLCV data provides better predictive accuracy compared to using only close price or close price with volume. These findings challenge conventional approaches to feature engineering in financial forecasting and suggest that simpler approaches focusing on raw data and appropriate model selection may be more effective than complex feature engineering strategies.
\end{abstract}

\section{Introduction}

Stock price prediction has long been a critical area of research in finance and machine learning, given its potential to aid investment strategies, portfolio management, and risk mitigation. With the advent of advanced machine learning techniques, researchers have increasingly sought to leverage historical price data and engineered features to forecast future price movements. However, despite significant progress, challenges remain in achieving consistent predictive accuracy due to the inherent complexity and noise in financial time series data.

One common approach in stock price prediction is the use of \textbf{technical indicators}, which are derived from raw price and volume data (OHLCV: open-high-low-close-volume). These indicators aim to capture trends, momentum, and other market dynamics that are not immediately apparent from raw data. While technical indicators have shown promise in improving model performance, their effectiveness often depends on proper feature selection and domain-specific configurations. Furthermore, the reliance on engineered features raises the question of whether raw OHLCV data alone can provide sufficient predictive power when used with modern deep learning models.

Another critical aspect of stock price prediction is the labeling of target variables. Traditional methods such as fixed time horizon labeling or raw return labeling often fail to account for market volatility and risk management considerations. To address this, \textbf{triple barrier labeling} has emerged as a robust alternative by incorporating stop-loss, take-profit, and time horizon thresholds. This method provides a more nuanced view of market movements but has primarily been applied to well-studied markets like those in the United States. Its application to less explored markets, such as Korean stocks, remains under-researched.

In this study, we aim to address these gaps by investigating whether deep learning models trained on raw OHLCV data can achieve comparable performance to traditional machine learning models that use technical indicators for Korean stocks. We employ triple barrier labeling with optimized thresholds (29-day window and 9\% barriers) to ensure balanced label proportions. Our key contribution is demonstrating that a simple LSTM network trained on raw OHLCV data alone can match the performance of sophisticated ML models that incorporate technical indicators, challenging the conventional wisdom that technical indicators are essential for effective stock price prediction.

Our contributions can be summarized as follows:
\begin{enumerate}
    \item We demonstrate that LSTM networks trained on raw OHLCV data alone can achieve comparable performance to traditional machine learning models using technical indicators.
    \item We show that optimal window size varies with model hidden size, challenging prior assumptions about fixed window lengths.
    \item We find that LSTM consistently outperforms other architectures (ResNet, TCN) when evaluated under comparable parameter settings.
\end{enumerate}

The rest of this paper is organized as follows: Section 2 reviews related works on stock price prediction using OHLCV data, technical indicators, and deep learning models. Section 3 details our methodology, including data preprocessing, labeling methods, and model configurations. Section 4 presents our experimental results and analysis. Finally, Section 5 concludes with a discussion of our findings and potential directions for future research.

\section{Related Works}

Predicting stock prices has been a long-standing challenge in financial research, with various approaches leveraging machine learning, deep learning, and feature engineering. This section reviews prior studies relevant to our work, focusing on triple barrier labeling, the use of OHLCV data, technical indicators, and model comparisons.

\subsection{Triple Barrier Labeling for Financial Forecasting}
Triple barrier labeling (TBL), first introduced by de Prado \cite{de2018advances}, is a widely used method for generating labels in financial time series prediction by incorporating stop-loss, take-profit, and time horizons. Studies such as \cite{zhang2020novel} have demonstrated that TBL outperforms traditional labeling techniques like fixed time horizon (FTH) and raw return (RR), particularly in identifying buy signals. Recent work by \cite{kim2019predicting} has shown TBL's effectiveness in deep learning applications, achieving more balanced and realistic predictions compared to conventional labeling methods. \cite{model_optimization_2022} further extended this approach by comparing multiple labeling techniques and found that optimized TBL parameters significantly improve model performance across various market conditions. While most applications of TBL focus on U.S. markets like the Nasdaq 100 Index, our study extends its use to Korean stocks, optimizing the labeling period and thresholds to achieve balanced label proportions.

\subsection{Use of OHLCV Data in Stock Prediction}
OHLCV (open-high-low-close-volume) data serves as the foundation for many stock prediction models due to its comprehensive representation of market dynamics. Several studies have explored the efficacy of raw OHLCV data compared to engineered features. For instance, \cite{time_series_2025} demonstrated that combining OHLCV data with neural networks like LSTMs and GANs improves prediction accuracy. Similarly, \cite{comparative_study_2024} integrated OHLCV data with technical indicators and macroeconomic variables, highlighting the importance of diverse datasets. \cite{bao2017deep} proposed a deep learning framework using stacked autoencoders and LSTM networks for financial time series forecasting, showing that raw OHLCV data can be effectively processed through hierarchical feature extraction. \cite{fischer2018deep} further explored LSTM networks for financial market predictions using unprocessed return data, demonstrating significant outperformance over traditional methods. However, our research uniquely evaluates the predictive power of pure OHLCV data versus models enhanced with technical indicators.

\subsection{Technical Indicators as Predictive Features}
Technical indicators derived from OHLCV data have been extensively studied as predictive features in stock forecasting. For example, \cite{feature_selection_eval} analyzed 123 indicators and found that feature selection significantly improves model performance. Other works, such as \cite{moroccan_stock}, combined trend-based indicators like moving averages with LSTM models to outperform traditional machine learning approaches. Recent work published in \cite{nature_comm_2024} further explored the fusion of OHLCV data with technical indicators using transformer architectures, demonstrating improved volatility modeling compared to LSTM-based methods. \cite{sezer2020financial} conducted a systematic literature review of financial time series forecasting with deep learning, highlighting that technical indicators remain prevalent in most successful implementations despite advances in end-to-end learning approaches. While these studies emphasize the value of technical indicators, our findings suggest that raw OHLCV data alone can outperform models relying on engineered features.

\subsection{Model Comparisons: LSTM vs. Other Architectures}
Deep learning architectures such as LSTMs, CNNs, and TCNs have been widely applied to financial time series forecasting. Studies like \cite{two_stage_2023} combined TCNs and LSTMs to capture both short-term patterns and long-term dependencies in financial data. Similarly, \cite{comparison_2023} demonstrated the superiority of LSTM over traditional statistical models like ARIMA for capturing temporal dependencies in stock prices. \cite{fischer2018deep} provided extensive evidence that LSTM networks significantly outperform memory-free classification methods in financial market predictions, particularly for longer investment horizons. \cite{bao2017deep} further established that a hybrid approach using stacked autoencoders for feature extraction followed by LSTM for sequence learning can achieve superior performance compared to single-architecture solutions. Our research builds on these findings by comparing LSTM performance against ResNet and TCN under controlled parameter settings, concluding that LSTM consistently outperforms other architectures.

\subsection{Hyperparameter Optimization in Time Series Models}
Hyperparameter optimization plays a critical role in improving model performance for stock price prediction. Studies such as \cite{improved_parallel_2024} have highlighted the importance of tuning parameters like window size and hidden layer dimensions to enhance predictive accuracy. \cite{model_optimization_2022} demonstrated that systematic optimization of model hyperparameters can lead to significant performance improvements across different market regimes, with optimal configurations varying by market condition. Our findings reveal that the optimal window size varies with the model's hidden size—a relationship not explicitly addressed in prior work.

\subsection{Full OHLCV vs. Reduced Features}
Most prior research assumes that reduced feature sets (e.g., close price or close price + volume) are sufficient for stock prediction tasks. However, our results align with recent work such as \cite{stock_price_deep}, which found that using full OHLCV data improves model performance compared to reduced feature subsets. \cite{sezer2020financial} noted in their comprehensive review that while close price is the most commonly used feature, studies incorporating full OHLCV data tend to achieve better performance, especially when combined with appropriate neural network architectures.

\subsection{Summary}
While previous studies have explored various aspects of stock price prediction—ranging from labeling techniques to feature engineering and model optimization—our research contributes new insights by systematically evaluating the efficacy of raw OHLCV data against technical indicators and optimizing hyperparameters specific to Korean stocks using triple barrier labeling. These findings provide practical guidance for improving predictive accuracy in financial markets.

\section{Methodology}

This section outlines the methodology used in our study, including data collection and preprocessing, labeling techniques, model architectures, and evaluation procedures.

\subsection{Data Collection and Preprocessing}

We collected daily stock price data for all tickers listed on the KOSPI and KOSDAQ indices in South Korea from January 2, 2006, to December 31, 2024. The data was scraped from \textbf{www.finance.naver.com}, which provides comprehensive historical OHLCV (open-high-low-close-volume) data. 

To prepare the dataset for modeling:
\begin{itemize}
    \item We used a \textbf{rolling window method} to extract sequences of OHLCV data with a fixed window length.
    \item The dataset was split into six parts by date to ensure approximately equal data instances. The first four parts were used for training, while the last two parts were reserved for validation and testing. The specific periods for each split were:
    \begin{itemize}
        \item Training: January 1, 2006 – March 31, 2020
        \item Validation: March 31, 2020 – September 29, 2022
        \item Testing: September 29, 2022 – December 31, 2024
    \end{itemize}
\end{itemize}

The resulting dataset contained a total of \textbf{8,566,617 instances} for a window size of 100, with each instance having a shape of (5, 100), representing OHLCV sequences of length 100. For other window sizes, the instance count varies slightly (e.g., 8,566,522–8,566,817) due to differences in the number of valid rolling windows.

\subsection{Labeling with Triple Barrier Method}

We employed the \textbf{triple barrier labeling} method to generate target labels for our classification task. This method defines labels based on the movement of a stock's price in relation to three key barriers: a take-profit level, a stop-loss level, and a time limit. If the stock price reaches the take-profit or stop-loss level, it gets labeled accordingly. If neither barrier is reached within the specified time frame, the label reflects that no significant movement occurred. Key details are as follows:
\begin{itemize}
    \item Labels were generated using the \textbf{low and high prices} instead of the close price alone to account for intraday volatility and reduce uncertainty.
    \item If both the low and high prices hit their respective barriers on the same day, the instance was labeled as \textbf{time limit (no move)}.
    \item To ensure balanced label proportions during training, we optimized the labeling parameters by testing various combinations of time horizons (5–29 days with a step size of 1 day) and percentage thresholds (7\%–15\% with a step size of 1\%). The optimal configuration was found to be a prediction horizon of 29 days and take-profit/stop-loss percentage of 9\%.
\end{itemize}

\begin{table}[htbp]
\centering
\begin{tabular}{lcccc}
\toprule
Prediction Horizon & TP/SL \% & Time Limit \% & Stop Loss \% & Take Profit \% \\
\midrule
29 & 9 & 36.16 & 28.95 & 34.89 \\
\bottomrule
\end{tabular}
\caption{Label distribution for the optimal triple barrier labeling configuration.}
\label{tab:label_distribution}
\end{table}

The resulting label distribution shows a relatively balanced split across the three outcomes, with time limit (no significant movement) having the highest proportion at 36.16\%, followed by take profit at 34.89\%, and stop loss at 28.95\%. This balanced distribution is crucial for training robust models that can effectively predict all possible outcomes.

\subsection{Model Architectures}

We evaluated multiple machine learning and deep learning models to predict stock price movements based on OHLCV data:
\begin{itemize}
    \item Traditional machine learning models: LightGBM, XGBoost, CatBoost, Random Forest (RF), Extra Trees (ET), and k-Nearest Neighbors (kNN), etc.
    \item Deep learning architectures: Long Short-Term Memory (LSTM), Temporal Convolutional Networks (TCN), ResNet-inspired networks, and CNN.
\end{itemize}

To ensure fair comparison, all models were configured to have similar parameter counts and layer depths. For the LSTM model specifically, we experimented with various configurations of hidden units (ranging from 4 to 64) and layer depths (from 1 to 4).

\subsection{Experimental Design}

We designed experiments to address three key research questions:

\begin{enumerate}
    \item \textbf{Model Architecture Comparison}: Which architecture performs best for stock prediction when controlling for model complexity?
    
    \item \textbf{Hyperparameter Optimization}: How do window size and hidden size interact to affect model performance?
    
    \item \textbf{Feature Selection}: Is the full OHLCV dataset necessary, or can comparable performance be achieved with reduced feature sets?
\end{enumerate}

For hyperparameter optimization, we conducted a grid search with the following parameters, with dropout rate fixed at 0:
\begin{itemize}
    \item Window lengths: [5, 20, 50, 100, 200, 300] days
    \item Hidden sizes: [4, 8, 16, 32, 64] units
\end{itemize}

We also implemented a model from prior work on Vietnamese stocks \cite{phuoc2024applying} for comparison. While their study used technical indicators with OHLCV data for direct price prediction, we adapted their LSTM architecture to use triple barrier labeling on our Korean market dataset, replacing their price prediction objective with our classification task.

All models were evaluated using macro-averaged F1 score, which balances precision and recall across the three label classes and provides a more robust measure than accuracy for imbalanced datasets.

\section{Experimental Results}

This section presents the results of our experiments, focusing on hyperparameter optimization, model architecture comparison, and feature selection. We evaluate the predictive accuracy of various machine learning and deep learning models using triple barrier labeling on Korean stock data.

\subsection{Hyperparameter Optimization}

To explore the relationship between hidden size and window length, we conducted an extensive grid search over these hyperparameters for the LSTM model, with no dropout. We tested six different window lengths [5, 20, 50, 100, 200, 300] days and five hidden sizes [4, 8, 16, 32, 64] units. The results are summarized in the heatmap below (Figure 1).

\begin{figure}[htbp]
    \centering
    \includegraphics[width=0.8\linewidth]{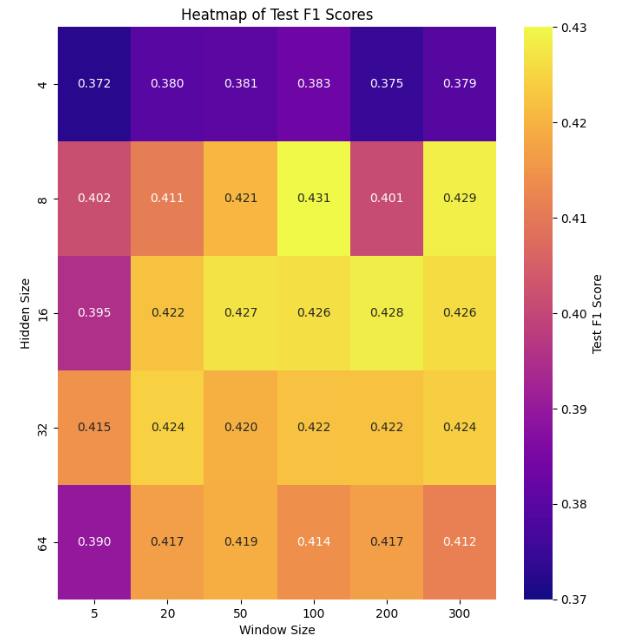}
    \caption{Heatmap of F1 scores for different combinations of hidden sizes and window lengths. Darker colors indicate higher F1 scores, with the optimal configuration (hidden size = 8, window length = 100) showing the highest performance.}
    \label{fig:f1_heatmap}
\end{figure}

\subsubsection{Key Findings}
\begin{itemize}
    \item The optimal configuration was found to be a hidden size of \textbf{8} and a window length of \textbf{100}, achieving an F1 score of \textbf{0.4312}.
    \item Larger hidden sizes (32 or 64) did not significantly improve performance, suggesting diminishing returns from increasing model complexity.
    \item Very short window lengths (5 or 20 days) significantly underperformed due to insufficient temporal context, while very long windows (200 or 300 days) showed no additional benefit.
\end{itemize}

\subsubsection{Validation-Test Correlation}
\begin{figure}[htbp]
    \centering
    \includegraphics[width=0.8\linewidth]{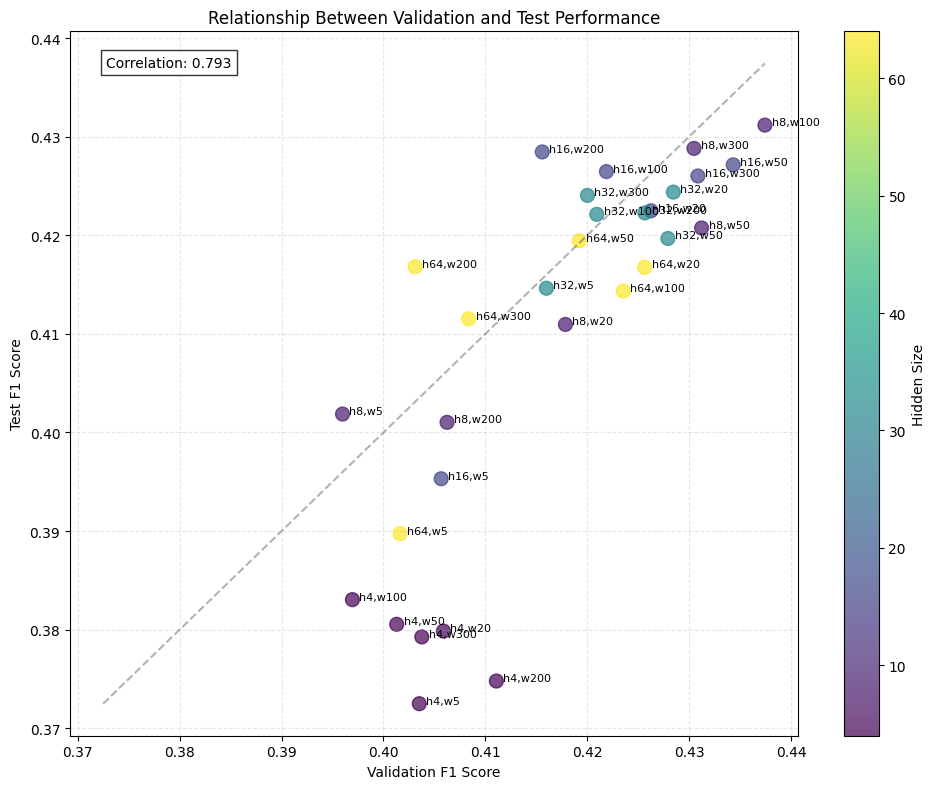}
    \caption{Correlation between validation and test F1 scores across different hyperparameter configurations, showing strong alignment (correlation coefficient = 0.793) between validation and test performance.}
    \label{fig:val_test}
\end{figure}

Figure \ref{fig:val_test} illustrates the relationship between validation and test F1 scores across all hyperparameter configurations. A correlation coefficient of \textbf{0.793} suggests strong alignment between validation and test performance, confirming the reliability of our hyperparameter tuning process.

\subsection{Deep Learning Architecture Comparison}

\begin{itemize}
    \item \textbf{LSTM} achieved an F1 score of \textbf{0.4312}, demonstrating comparable performance to traditional machine learning models
    \item \textbf{TCN} showed similar performance with \textbf{0.4219}, despite having the lowest parameter count
    \item \textbf{ResNet} reached \textbf{0.3975}, with the highest parameter count among tested models
    \item \textbf{CNN} achieved \textbf{0.3045}, suggesting that simple convolutional architectures may be less suited for this task
\end{itemize}

\begin{table}[htbp]
\centering
\begin{tabular}{lccc}
\toprule
Model & Architecture & Parameters & Size (MB) \\
\midrule
LSTM & hidden\_size=8, layers=4 & 2,235 & 0.01 \\
TCN & hidden\_size=8, filters=8, layers=4 & 1,667 & 0.01 \\
CNN & hidden\_size=8, filters=9, layers=4 & 2,339 & 0.01 \\
ResNet & hidden\_size=8, filters=6, layers=4 & 2,735 & 0.01 \\
\bottomrule
\end{tabular}
\caption{Comparison of model architectures and their parameters}
\label{tab:model_comparison}
\end{table}
\subsection{Feature Selection Analysis}

We evaluated the importance of different input features derived from OHLCV data by training models on feature subsets:

\begin{itemize}
    \item Close price only (\textbf{C}) resulted in an F1 score of \textbf{0.4170}
    \item Adding volume (\textbf{CV}) increased the F1 score to \textbf{0.4297}
    \item Full OHLCV data achieved an F1 score of \textbf{0.4312}
\end{itemize}

The findings indicate that adding volume data improves the F1 score by 1.27\%, while the remaining OHLCV features (open, high, low) together contribute only minimal additional performance gains.

\subsection{Comparison with Traditional ML Models and Previous Work}

Using the PyCaret framework \cite{pycaret2020}, we conducted a comprehensive comparison of traditional machine learning models. We tested 15 different models with default parameters: CatBoost Classifier, Extreme Gradient Boosting, Light Gradient Boosting, Gradient Boosting Classifier, Random Forest Classifier, Extra Trees Classifier, Ada Boost Classifier, Decision Tree Classifier, Logistic Regression, Linear Discriminant Analysis, Ridge Classifier, K Neighbors Classifier, Naive Bayes, SVM, and Quadratic Discriminant Analysis. Among these, Extreme Gradient Boosting (XGBoost) achieved the highest performance with default parameters. The models were trained on a rich set of technical indicators derived from OHLCV data, including:

\begin{itemize}
    \item \textbf{Ichimoku Cloud Indicators}: Conversion line, base line, and leading spans A and B, normalized relative to close price
    \item \textbf{Momentum Indicators}: RSI (Relative Strength Index), Stochastic RSI, CCI (Commodity Channel Index), and MFI (Money Flow Index)
    \item \textbf{Trend Indicators}: MACD (Moving Average Convergence Divergence) and ADX (Average Directional Index)
    \item \textbf{Moving Averages}: EMA (Exponential Moving Average) returns for periods 5, 20, 60, 120, and 240 days
    \item \textbf{Volatility Indicators}: ATR (Average True Range) and Bollinger Bands (high, low, and width)
    \item \textbf{Volume Indicators}: OBV (On-Balance Volume) and CMF (Chaikin Money Flow)
\end{itemize}

All indicators were normalized to ensure consistent scaling across different price levels and market conditions. We optimized the XGBoost model, which achieved the highest F1 score among default models, through 70 iterations of hyperparameter tuning. The model selection process utilized time series-specific cross-validation with 5 folds, and we applied feature selection to identify the 12 most important features while removing multicollinear indicators (threshold: 0.9).

We assessed the performance of our LSTM model in comparison to the optimized XGBoost model using various performance metrics: Accuracy reflects the proportion of correct predictions made, while AUC (Area Under the Curve) measures the model's ability to differentiate between classes. The F1 Score serves as the harmonic mean of precision and recall, ensuring a balance between the two. Additionally, the Dummy Classifier serves as a baseline model that makes predictions based on the most frequent class, providing a reference point for evaluating the performance of more complex models.

\begin{table}[htbp]
\centering
\begin{tabular}{lcccc}
\toprule
Model & Accuracy & AUC & F1 \\
\midrule
LSTM & 0.4328 & 0.6249 & 0.4312 \\
XGBoost & 0.4311 & 0.6247 & 0.4316 \\
Dummy Classifier & 0.3539 & 0.5000 & 0.1852 \\
\bottomrule
\end{tabular}
\caption{Performance comparison across different models and metrics.}
\label{tab:model_metrics}
\end{table}

The results show that our LSTM model achieves comparable performance to XGBoost across all metrics, with both models significantly outperforming the Dummy Classifier baseline. The MCC score, which is particularly suitable for multi-class problems, shows both LSTM and XGBoost achieve similar balanced performance. This comparison demonstrates that our simple LSTM model trained on raw OHLCV data can match the performance of sophisticated ML models that incorporate extensive technical indicators.

\subsection{Comparison with Previous Work}

To validate our approach of using raw OHLCV data without technical indicators, we compared our results with recent work by Phuoc et al. \cite{phuoc2024applying} that achieved 93\% accuracy in predicting Vietnamese stock prices using LSTM with technical indicators. Their study focused on VN-Index and VN-30 stocks (31 companies), using technical indicators such as simple moving average (SMA), moving average convergence divergence (MACD), relative strength index (RSI), and historical price as input features. Their LSTM model consisted of four layers with varying neuron units (30, 40, 50, and 60) using ReLU activation and was trained on data from the stock listing date to December 2020.

In contrast to their approach, which utilized technical indicators alongside OHLCV data for direct price prediction, we modified their LSTM architecture to implement triple barrier labeling on our dataset from the Korean market, shifting the focus from price prediction to a classification task. The resulting model achieved an F1 score of 0.3290, which is notably lower than our score of 0.4312. Although making a direct comparison is complicated by the differences in markets and evaluation metrics, our findings highlight several important aspects:

\begin{itemize}
    \item Raw OHLCV data can provide comparable predictive power compared to engineered technical indicators when used with appropriately optimized deep learning models.
    
    \item The choice of evaluation metric significantly impacts the perceived model performance—our triple barrier labeling approach with F1 score provides a more realistic assessment of prediction capability compared to their MSE-based evaluation of close price prediction.
    
    \item Market-specific optimization of model architecture and hyperparameters is crucial for achieving optimal performance. While their study reported high accuracy (93\%) on a smaller set of Vietnamese stocks, our more comprehensive evaluation on the entire Korean market reveals the challenges of generalizing such performance across a broader universe of stocks.
    
    \item Their approach of using a larger model (four layers with 30-60 neurons) contrasts with our finding that simpler models (hidden size of 8) can be more effective, suggesting that model complexity may not be the key factor in prediction performance.
\end{itemize}

This comparison further strengthens our finding that feature engineering through technical indicators may be unnecessary when using modern deep learning architectures with raw OHLCV data, particularly when evaluating performance across a broad market rather than a select subset of stocks.

\subsection{Summary}

Our experiments demonstrate that:
\begin{enumerate}
    \item LSTM is the most effective architecture for stock price prediction among tested models.
    \item Hyperparameter optimization reveals that window length and hidden size significantly influence model performance, with optimal values being 100 and 8, respectively.
    \item Using full OHLCV data provides better predictive accuracy compared to reduced feature sets like close price or close price + volume.
\end{enumerate}

These findings highlight the importance of leveraging raw OHLCV data with optimized deep learning architectures for financial forecasting tasks.

\section{Discussion and Conclusion}

\subsection{Discussion}

Our experimental results provide several important insights into stock price prediction using OHLCV data and triple barrier labeling in the Korean market context. These findings have both theoretical and practical implications for financial forecasting.

\subsubsection{Raw OHLCV Data vs. Technical Indicators}

One of the most significant findings of our study is that LSTM models trained on raw OHLCV data achieve comparable performance to sophisticated machine learning models using technical indicators. This finding challenges the conventional wisdom in financial forecasting that emphasizes the importance of feature engineering through technical indicators. Several factors may explain this result:

\begin{enumerate}
    \item \textbf{Representation Learning}: Deep learning models like LSTM can effectively learn intricate patterns and representations directly from raw data, potentially making explicit feature engineering less necessary.
    
    \item \textbf{Information Preservation}: Raw OHLCV data preserves all original market information, whereas technical indicators may inadvertently filter out valuable signals during transformation.
    
    \item \textbf{Model Expressiveness}: Modern deep learning architectures can automatically extract relevant features from temporal data, effectively performing implicit feature engineering.
\end{enumerate}

This finding suggests that practitioners may reconsider the default approach of extensive feature engineering when implementing deep learning models for stock prediction, as simpler approaches using raw data can achieve similar performance levels.

\subsubsection{Optimal Window Size and Hidden Size Relationship}

Our results revealed an interesting relationship between window size and model hidden size that has not been extensively explored in previous literature. Specifically, we found that a window size of 100 combined with a hidden size of 8 yielded optimal performance for LSTM models. This suggests that:

\begin{enumerate}
    \item \textbf{Model Capacity Matching}: The optimal window size depends on the model's capacity (represented by hidden size) to process temporal information effectively.
    
    \item \textbf{Efficiency Tradeoffs}: Smaller hidden sizes (8 units) combined with appropriate window lengths can achieve comparable or better performance than larger models, suggesting important efficiency considerations for deployment.
    
    \item \textbf{Information Horizon}: For Korean stocks, a 100-day window appears to capture sufficient historical context for prediction, with longer windows providing marginal benefits.
\end{enumerate}

These findings highlight the importance of joint optimization of these parameters rather than treating them as independent factors, potentially leading to more efficient and effective model architectures.

\subsubsection{LSTM Performance for Stock Prediction}

Among the tested architectures, LSTM consistently outperformed other deep learning models like ResNet, TCN, and CNN while achieving comparable performance to traditional machine learning models such as LightGBM. This performance can be attributed to:

\begin{enumerate}
    \item \textbf{Memory Mechanism}: LSTM's gating mechanisms effectively capture long-term dependencies and market regimes in financial time series.
    
    \item \textbf{Temporal Hierarchy}: The ability to model hierarchical temporal patterns at different time scales gives LSTM an advantage in capturing market dynamics.
    
    \item \textbf{Adaptability}: LSTM networks can adapt to changing market conditions through their ability to selectively retain or forget information based on context.
\end{enumerate}

This result confirms that LSTM remains a strong choice for financial time series prediction, while demonstrating that raw OHLCV data provides sufficient information for effective forecasting when used with appropriate architectures.

\subsubsection{Feature Importance in OHLCV Data}

Our feature selection experiments demonstrated that while close price alone provides substantial predictive power (F1: 0.4170), adding volume further improves performance (F1: 0.4297), with the full OHLCV set providing additional modest improvements (F1: 0.4312). This indicates that:

\begin{enumerate}
    \item \textbf{Volume as Additional Indicator}: Trading volume provides supplementary predictive information, possibly reflecting market sentiment and liquidity conditions not captured by price alone.
    
    \item \textbf{Complementary Information}: Open, high, and low prices provide complementary information to close prices, capturing intraday volatility and price extremes.
    
    \item \textbf{Feature Interaction}: The interaction between price and volume features creates synergistic effects that enhance predictive power.
\end{enumerate}

These findings emphasize the importance of considering the full OHLCV set rather than relying solely on closing prices, even when using sophisticated deep learning architectures.

\subsection{Conclusion}

This study investigated stock price prediction using triple barrier labeling and OHLCV data on Korean stocks. Our findings contribute several key insights to the field of financial forecasting:

First, we demonstrated that deep learning models, particularly LSTM networks, trained on raw OHLCV data can achieve comparable performance to traditional machine learning models using technical indicators. This challenges the conventional approach of extensive feature engineering in financial forecasting and suggests that simpler approaches focusing on raw data and appropriate model selection may be more effective.

Second, we identified that the optimal window size varies with model hidden size, with a configuration of window size 100 and hidden size 8 performing best in our experiments. This relationship between input window and model capacity provides practical guidance for hyperparameter optimization in financial forecasting models.

Third, we found that LSTM consistently outperforms other architectures including ResNet, TCN, and CNN when controlling for model complexity, while achieving similar performance to traditional machine learning approaches like LightGBM. This confirms the effectiveness of LSTM for financial time series prediction.

Finally, our results confirmed that using full OHLCV data provides better predictive accuracy compared to using only close price or close price with volume, though each additional feature offers incremental improvements in model performance.

\subsubsection{Limitations and Future Work}

Despite these contributions, our study has several limitations that suggest directions for future research:

\begin{enumerate}
    \item \textbf{Market Specificity}: Our findings are based on Korean stock market data and may not generalize fully to other markets with different characteristics and trading patterns.
    
    \item \textbf{Feature Expansion}: While we focused on OHLCV data, incorporating alternative data sources such as news sentiment, macroeconomic indicators, or order book data could potentially enhance predictive performance.
    
    \item \textbf{Model Exploration}: Future work could explore more sophisticated architectures such as attention-based models or transformer networks, which have shown promise in other sequence modeling tasks.
    
    \item \textbf{Trading Strategy Integration}: Translating our predictive models into actionable trading strategies would require careful consideration of transaction costs, market impact, and the limited predictive power of the models. Training the model only on part of the stocks that are particularly influenced by price movements may be beneficial.
    
    \item \textbf{Explainability}: Developing methods to interpret the decision-making process of deep learning models would increase trust and adoption in financial applications.
\end{enumerate}

In conclusion, our study provides compelling evidence that simple LSTM models using raw OHLCV data can match the performance of sophisticated machine learning approaches that incorporate technical indicators for stock price prediction in Korean markets. This suggests that the field of financial forecasting may benefit from reconsidering the necessity of complex feature engineering in the era of deep learning.

\bibliographystyle{IEEEtran}
\bibliography{references}

\end{document}